\documentstyle[aas2pp4]{article}

\def\gs{{_>\atop^{\sim}}}

\include{AAS_defs}

\begin{document}

%\sffamily
%\Large

%%%%%%%%%%%%%%%%%%%%%%%%%%%%% TITLE PAGE %%%%%%%%%%%%%%%%%%%%%%%%%%%%%

\title{Deriving the Quasar Luminosity Function from
 Accretion Disk Instabilities}

\author{Aneta Siemiginowska \& Martin Elvis}

\bigskip

\affil{Harvard-Smithsonian Center for Astrophysics \\
60 Garden Street, MS-70,  Cambridge, MA~02138, USA \\
 asiemiginowska@cfa.harvard.edu}

\begin{center}
\bf
\today

\end{center}

%%***************************************************************************
%%  ABSTRACT
%%***************************************************************************

\begin{abstract}
We have derived the quasar luminosity function assuming that the
quasar activity is driven by a thermal-viscous unstable accretion disk
around a supermassive black hole. The instabilities produce large
amplitude, long-term variability of a single source. We take a light
curve of a single source and calculate the luminosity function, from
the function of time it spends at each luminosity.  Convolving this
with an assumed mass distribution we fit well the observed optical
luminosity function of quasars at four redshifts.  As a result we
obtain the evolution of the mass distribution between redshifts 2.5
and 0.5.

The main conclusions are following:
1) The quasar long-term variability due to the disk thermal-viscous
instabilities provides a natural explanation for the observed quasar
luminosity function.
2) The peak of the mass function evolves towards lower black hole masses
at lower redshifts by a factor $\sim 10$.
3) High mass sources die subsequently when redshift gets smaller.
4) The
number of high mass sources declines rapidly, and so low mass sources
become dominant at lower redshift.
5) The periodic outbursts of activity appear as long as the matter is
supplied to the accretion disk.
6) Since the time-averaged accretion rate is low, the remnant sources
(or sources in the low activity phase) do not grow to very massive
black holes.
7) A continuous fuel supply at a relatively low accretion rate ( $\sim
0.01 - 0.1 \dot M_{Edd}$) for each single source is required over the
lifetime of the entire quasar population.
\end{abstract}


\section{INTRODUCTION}

The quasar luminosity function has been studied for the last 30 years
and is observationally now quite well determined as a function of
redshift for $z<4$ (e.g. Boyle et al, \markcite{b2} 1991).  However,
there have been few attempts to derive the luminosity function from
physical models of the quasar power engine.  There are three possible
phenomenological scenarios (Cavaliere \& Padovani, \markcite{c5} 1988):
long-lived objects, recurrent objects (possibly related to galaxy
mergers) and a single short event over the whole host galaxy
life-time.  Continuous models imply masses for the remnant black holes
that are too large, and accretion rates that are too low (Cavaliere et
al \markcite{c3} 1983; Cavaliere \& Szalay  \markcite{c4} 1986;
Cavaliere \& Padovani \markcite{c5} 1988,\markcite{c6} 1989; Caditz,
Petrosian \& Wandel, \markcite{c1} 1991).

Short-lived models have been studied more recently.  Haehnelt \& Rees
\markcite{h1} (1993) assumed that new quasars were born at successive
epochs with a short active phase followed by a rapid exponential
fading due to exhaustion of fuel.  They used the Cold Dark Matter
formalism (Press \& Schechter, \markcite{p1} 1974) to estimate the
number of newly forming dark matter halos at different cosmic epochs.
Small \& Blandford \markcite{s4} (1992) suggested a scenario involving
a mixture of continuous and recurrent activity. They assumed that
newly formed sources achieve the Eddington luminosity quickly, such
that the accretion rate is limited by radiation pressure. The break in
the luminosity function is related to the boundary between the
continuous and intermittent accretion phases originating in the amount
of fuel supply to the black hole.

However, none of these models relate directly to the physical
processes responsible for powering a quasar.  They simply invoke
sources that emit at the Eddington luminosity for a certain time and
then fade below an observational threshold.  Here we describe a
scenario which for the first time derives the luminosity function from
a specific physical process.

The time evolution of an accretion disk around a supermassive black
hole (the main components of the standard quasar paradigm), exhibits
large variations on long timescales due to thermal-viscous
instabilities (Siemiginowska, Czerny \& Kostyunin \markcite{s2} 1996,
hereafter SCK96, Mineshige \& Shields \markcite{m5} 1990).  Depending
on the assumed disk model, variations of up to a factor $\sim 10^4$
can be produced on timescales of 10$^4-10^6$ years.  Here, we assume
that all quasars are subject to this variability. We then take the
light curve of a single source and calculate the luminosity function
of a population of identical sources from the fraction of time it
spends at each luminosity.  Convolving this with an assumed mass
distribution we fit the observed quasar luminosity function at four
redshifts.  As a result we obtain the evolution of the mass
distribution between redshifts 2.5 and 0.5.

%Thermal-viscous instabilities are discussed briefly in the Section
%II. Section III shows how the luminosity function was derived based on
%the assumed light curve and mass function. Sections IV contains
%the discussion.

\section{EVOLUTION OF AN ACCRETION DISK}

Accretion onto a supermassive black hole is the leading model for
powering quasars (e.g. Rees \markcite{r1} 1984). The accretion process
is frequently described by the model of a stationary thin disk
(Lynden-Bell \markcite{l2} 1969, Shakura \& Sunyaev \markcite{s1}
1973).  However, there are both observational and theoretical
arguments indicating that time-dependent effects in accretion process
are of extreme importance.  Observationally, the evidence for global
evolutionary effects is compelling in accretion disks around Galactic
X-ray sources. Outbursts (by factors $>10^4$) of Cataclysmic Variables
or X-ray novae last for weeks or months and happen every few months to
years.  The outbursts are essentially caused by the disk thermal
instability in the partial ionization zone (Meyer \& Meyer-Hoffmeister
\markcite{m4} 1982, Smak \markcite{s3}  1982, see also Cannizzo
\markcite{c2} 1993 for review).  There is a strong similarity between
Galactic X-ray sources and AGN both in spectral behavior and overall
variability (Fiore \& Elvis \markcite{f2} 1994, Tanaka
\& Lewin \markcite{t1} 1995) which leads us to expect similar accretion disk
behavior in AGN. However, as the characteristic timescales are roughly
proportional to the central mass, the expected variability takes
thousands to millions of years in AGN. Since these timescales are not
directly observable, these changes have been considered little more
than a curiosity in AGN.

Theoretically, accretion disks around the massive black holes in AGN
are expected to have a partial ionization zone, as in Galactic
binaries, and therefore to be subject to the same instability (Lin \&
Shields \markcite{l1} 1986, Clarke \markcite{c7} 1989,
%Siemiginowska, Czerny \& Kostyunin
SCK96\markcite{s2}).
% \markcite{s2} 1996).
Current models of the time evolution of accretion disks in AGN have
confirmed the presence of disk eruptions (Clarke \& Shields
\markcite{c8} 1989, Mineshige \& Shields \markcite{m5} 1990,
%Siemiginowska, Czerny \& Kostyunin \markcite{s2}
SCK96\markcite{s2}).

SCK96\markcite{s2}
%Siemiginowska, Czerny \& Kostyunin
%\markcite{s2} (1996)
considered a geometrically thin Keplerian accretion disk around a
supermassive black hole and assumed that the viscosity scales with the
gas pressure ($\tau_{r\phi} = \alpha P_{gas}$).
They found that, depending on the viscosity, the instability can
either develop only in a narrow unstable zone, or can propagate over
the entire disk resulting in large amplitude optical/ultraviolet
outbursts ($\sim 10^4$) (see Fig.~1a). The calculation of these light
curves is at present computationally demanding (SCK96\markcite{s2}).

\section{FROM LUMINOSITY VARIATIONS  TO THE LUMINOSITY FUNCTION}

\subsection{A single mass, single accretion rate  population.}

The luminosity function of a population of quasars with the same mass
and accretion rate is given simply by the product of a fraction of
time one source spends in each luminosity bin and their space density.
In Fig.1b we show the fraction of time a source emits at each luminosity,
relative to the Eddington luminosity, for the light curve shown in
Fig.1a. The luminosity range is between $10^{-4} L_{Edd}$ and
$L_{Edd}$. The shape of the function reflects the fact that
the amplitude of each outburst is not constant and the variability is
not precisely periodic. The details of each outburst and the overall
variability characteristics depend on the physics of the accretion
disk and the assumptions of the model. These details average
over many outbursts (usually a few hundred over 10$^8-10^9$ years).

\begin{figure}[h]
\centering
\plotone{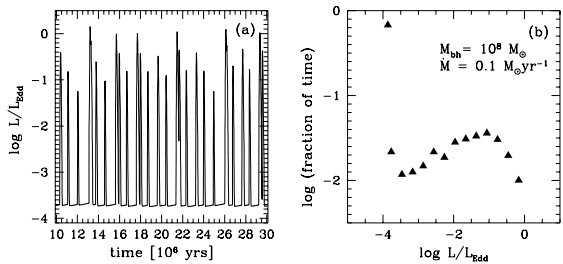}
\figcaption[fig1.eps]{
a) Luminosity variations due to the disk instabilities around a black
hole of $10^8 M_{\odot}$, when the accretion rate is
0.1~M$_{\odot}$~yr$^{-1}$ and the viscosity parameter is different in
the high and low states: $\alpha_{hot}=0.1$ and
$\alpha_{cold}=0.025$.
b) Fraction of the time the source emits at a given luminosity
for $10^8 M_{\odot}$ and 0.1$\dot M_{\odot}$yr$^{-1}$ accretion rate. The
luminosity is given in the Eddington luminosity units. \label{fig1}}
\end{figure}

A single source will spend about $\sim 75\% $ of its life in
quiescence ($L< 0.001 L_{Edd}$) and about $\sim 25\%$ in an active
state, with $\sim 10\%$ in a high state ($L> 0.1 L_{Edd}$).  Likewise
in a population $\sim$10\% of sources will be in the high state,
$\sim$25\% will be active and $\sim$75\% in quiescence at any given
time.

There are two characteristic transition points in the function shown
in Fig.1b: a broad maximum at $\sim 0.1 L_{Edd}$ and a minimum at
$\sim 0.001 L_{Edd}$. When we construct a luminosity function for a
realistic population these features will be modified by the
distribution of accretion rates and masses.  A range of accretion
rates affects the low luminosity part of the curve by smoothing at the
minimum.  The maximum at $\sim 0.1 L_{Edd}$ is caused by the fraction
of outburst amplitudes reaching close to the Eddington
luminosities. The maximum is thus smoothed by the distribution of
black hole masses.

\subsection{Fit to the Observed Luminosity Function.}

The luminosity in Fig.1b. is expressed in terms of the Eddington
luminosity, in order to make the function independent of the central
mass (see Fig.2a ``1-mass'' luminosity function). Thus for a given
distribution of black hole masses we can calculate the quasar
luminosity function.
The luminosity function is defined as:

	$$ \Phi (L,z)  = \int \Phi (L,z,M) N(M,z) dM  $$

\noindent
where $\Phi (L,z,M)$ describes which central mass contributes to a
given luminosity bin at a given redshift and $N(M,z)$ represents a
number of sources with a given central mass at a given redshift. The
mass density function ($N(M,z) M$) can be derived, with assumptions,
from cosmological models and theoretical models on the formation of
the structures in the universe (Heahnelt \& Rees \markcite{h1} 1993,
Small \& Blandford \markcite{s4} 1992). Here we do not consider any
particular model for the formation of the black holes, galaxies and
quasars.  Instead we assume, arbitrarily that $N(M,z)$ can be
represented by a simple parabola. We then fit the observed luminosity
function from Boyle et al \markcite{b2} (1991) at different epochs
varying the peak, position and width of the parabola. The mass
function is convolved with the single mass luminosity function
(Fig.1b). We held $\dot M$ constant, and so did not change the minimum
in Fig.1b. However, this minimum is not within the range covered by
the Boyle et al \markcite{b2} (1991) data.

\begin{figure}[h]
\centering
\plotone{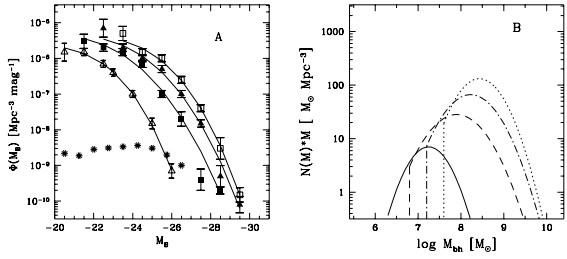}
\figcaption[fig2.eps]{
 (a) Observed luminosity function (Boyle et al 1991) at four redshifts
indicated by points with error bars: $0.25<z<0.75$ - open triangles,
$0.75<z<1.25$ - filled squares, $1.25<z<2.00$ - filled triangles,
$2.00<z<2.90$ - empty squares. The predicted single mass luminosity
function is indicated with the stars. The model luminosity function is
plotted with the solid line.  (b) The predicted black hole mass
density function at each epoch obtained from the fit to the observed
luminosity function: $0.25<z<0.75$ - solid line; $0.75<z<1.25$ -
dashed line $1.25<z<2.00$ - dashed-dot line $2.00<z<2.90$ - dotted
line.  The truncation at the mass required by the lowest luminosity
data point is indicated by the straight lines at the lowest
mass point for each parabola. \label{fig2}}
\end{figure}

We were able to obtain good fits (Table 1) to the luminosity function
at three of four redshifts as shown in Fig.~2a.  The black hole mass
density function required by the fit is plotted in Fig.~2b and the
parameters of the fit are given in the Table 1.  The poor fit at
redshift $\sim$1 is due to the highest luminosity point which seems to
require a kink in the luminosity function. We fit the data excluding
this point and obtain a good fit (Table 1).  The mass density
peaks at lower mass in this case, since the luminosity function ends
at lower luminosity.

\begin{table}
\begin{center}
\tablecaption{Results of the Modeling \label{tab1}}
\begin{tabular}{cccc}
& & &   \\
redshift & Log M$_{peak}$ & log N(M$_{peak}$) & $\chi ^2$ \\
 z &  & &   \\
\tableline
 &&& \\
 2.0 - 2.9  & 8.1 &  -5.30 &   2.78 \\   %  7.20000   -5.30&  7.7 & 131.8 &
 1.25 - 2.0 & 7.7 &  -5.33 &   4.98 \\   %  6.90000   -5.33& 7.4 &  66.1 &
 0.7 - 1.25 & 7.3 &  -5.34 &   12.79 \\  %  6.50000   -5.34& 7.1 & 28.2 &
 	    & 6.9 & - 5.38 &	3.48 \tablenotemark{a} \\
 0.3 - 0.7  & 6.9 &  -5.52 &   1.18 \\	%  6.20000   -5.52 & 6.6 &   6.9 &
 & & &  \\
\end{tabular}
\end{center}
\tablenotetext{a}{
excluding the highest luminosity data point at redshifts  $0.7<z<1.25$}
\end{table}

Only active sources contribute to the black hole mass density
function.  The maximum indicates which central black hole mass
dominates the population at each redshift.  This peak mass declines
from $\sim 2\times 10^7 M_{\odot}$ at $z
\sim 2.5$, to $\sim 2\times 10^6 M_{\odot}$ at $z \sim 0.5$ with
most of the change occuring between $z \sim 1.75$ and $z \sim 1$
(Table 1).  The mass density of high mass sources gets smaller rapidly
(e.g. a factor $\sim 100$ at $ M = 10^9 M_{\odot}$) with lower
redshift, while the peak mass density declines by less than a
factor 2. The relative constancy of this peak implies that the {\em
number} of sources at the peak remains constant between $z=2.5$ and
$z=1$, and then decreases by a factor of $\sim 2$ between redshift
$z=1$ and $z=0.5$. The low mass end of the distribution is not
constrained by the data and so we truncate the functions at the mass
where the Eddington limit gives the lowest observed luminosity. This
lowest luminosity point of the observed luminosity function at $z=2.5$
(L$_{Edd} \sim 8 \times 10^{44}$~ergs~s$^{-1}$) gives the limit of
$6.3 \times 10^7 M_{\odot}$.  Even if less massive sources are
present in the population we cannot see them.

\section{DISCUSSION}

We have shown that the thermal-viscous instability provides a natural
mechanism to generate the quasar luminosity function.  We were able to
fit the observed luminosity function and estimate the parameters of
the mass density function, independent of cosmological models.
%The
%peak of the mass distribution moves towards lower masses when the
%redshift gets smaller, and
The overall shape of the mass density function and the evolution of
the peak of the mass distribution towards lower masses with lower
redshift are similar to the results obtained by previous studies
(Haehnelt \& Rees\markcite{h1} 1993, Small \& Blandford\markcite{s4}
1992).  The 30\% ``on'' fraction in these models is also comparable
with the fraction of active time input light curve (Fig.1b).  This is
not too surprising because the same observational luminosity function
was used in all the studies.

The problem of whether the low mass sources are present at high
redshift or are born subsequently remains unsolved. In our scenario
this question could be answered by extending the observed luminosity
function at z$\gs 1$ fainter by $\Delta m \sim 3$. If the fitted mass
functions match the low $z$ mass density functions these would suggest
that all quasars are born at the same time and the high mass ones
``burn out'' much more quickly.  This scenario, in which single mass
density function declines more rapidly at high luminosities, is
strikingly different from the conventional ``pure luminosity
evolution'' that is used to describe the apparent fading of the whole
observed luminosity function to lower $z$. It reminds us of the
warning by Green \markcite{g1} (1985) against interpreting
phenomenological descriptions as physically meaningful.

Looking at the sources in the present epoch should provide the
information on the lowest luminosity end of the distribution together
with the contribution of the massive sources.  However, the luminosity
of the host galaxy becomes comparable to the nuclear luminosity for
low mass black hole and it is hard to observed the nucleus of a normal
galaxy even if it contains an accretion disk in the active state ($L
\sim L_{Edd}$).  On the other hand Seyfert nuclei are found in
$\sim 10\%$ of galaxies, consistent with the high state fraction from
SCK96.\markcite{s2} The problem is how can we see the low mass sources
in quiescence.

The high luminosity end of the luminosity function accounts for all
the high mass sources which are active at each epoch. The number of
these sources gets smaller with redshift. This decrease is often
supposed to relate to the limited fuel supply and the mechanisms of
transfer of the matter into the disk. We note though that the location
and the size of the ionization zone depends on the accretion rate onto
the outer edge of the disk
%Siemiginowska, Czerny \& Kostyunin, \markcite{s2} 1996;
(SCK96,\markcite{s2} Clarke \& Shields \markcite{c8}1989). For high
accretion rates this zone moves towards outer regions of accretion
disk.  In the case of the high mass black holes the ionization zone
can be pushed out to the self-gravitating regions of the disk and the
instability will not develop. The source then remains in the active
state until the fuel supply is exhausted, and then dies. How the
location of the ionization zone affects the global evolution of the
population requires further study.

Another consideration that could lead to the more rapid demise of high
mass quasars is that massive sources require more fuel than the low
mass sources to emit at a given $L/L_{Edd}$. Only $\sim
0.027M_{\odot}$yr$^{-1}$ is needed to power a $10^6 M_{\odot}$ black
hole at 0.1~$\dot M_{Edd}$, while a $10^9 M_{\odot}$ black hole
requires accretion rates of order 2.7$M_{\odot}$yr$^{-1}$.  Recent
studies of quasar host galaxies, (at z$< 0.3$), show that the most
luminous quasars reside in the most massive galaxies, while lower
luminosity quasars can be found in any type of a galaxy (McLeod \&
Rieke\markcite{m2} 1995a,\markcite{m3} 1995b; Bahcall et
al\markcite{b1} 1996). Based on the HST observations of 61 elliptical
galaxies Faber et al\markcite{f1} (1996) conclude that about $\sim
1\%$ of the galaxy mass is contained within a central core of few
parsecs.  This means that for a typical $\sim 10^{11} - 10^{12}
M_{\odot}$ galaxy, there is about $10^9-10^{10} M_{\odot}$ available
to feed a black hole. While at $10^6 M_{\odot}$ it would last for
10$^{11}-10^{12}$ years at $10^9 M_{\odot}$ it would last for
10$^9$~years  at 0.1~$\dot M_{Edd}$.

A third possibility is that the, unknown, mechanism responsible
for transferring the matter towards the central potential well and
into the outer parts of accretion disks becomes s rapidly less
efficient in massive systems, so they would systematically die young.

Recently Yi \markcite{y1} (1996) considered the cosmological evolution
of quasars assuming that advection becomes important for accretion
rates below 0.01$L_{Edd}$ accretion rate.  The theoretical
and observational studies of the X-ray transients suggest that
advection is important in quiescence below a critical accretion rate
(Narayan \& Yi \markcite{n1} 1994, Narayan et al\markcite{n2}
1996). Advection has not been included in our accretion disk model. It
will modify the low luminosity part of the light curve in Fig.1a and
influence the quasar evolution. We shall include the advection in our
further studies, since the quasar remains in quiescence for $\sim
75\%$ of its life.

In previous studies the sources making up the luminosity function were
assumed to begin by emitting at their Eddington luminosities and
steadily becoming fainter with time.  This does not apply in our
model.  The stationary accretion rate onto the outer edge of the disk
can be much lower than the Eddington limit. This prevents accumulation
of a large mass in the center and removes the problem of creating
overly massive quasars remnants.

Small \& Blandford \markcite{s3} (1992) suggested two phases of the
quasar activity, which are related to the accretion rate.  Just after
a black hole is born the matter is supplied at super-Eddington rates,
but the actual accretion onto the black hole is limited by the
radiation pressure. The black hole accretes continuously at the
Eddington rate until the fuel supply gets lower and then the accretion
is intermittent. The intermittent activity can be related to the
active state of the disk in our scenario.

The model we use to produce the quasar luminosity function works for
the optical/ultraviolet bands. The radio and X-ray luminosity function
show similar form and evolution (Maccacaro et al. \markcite{m1} 1992,
Della Ceca et al \markcite{d1} 1994). Physically the radio
and X-ray luminosities must then be a result of the accretion disk state.

\acknowledgments

We thank Bo\.zena Czerny, Andrzej Soltan, Tom Aldcroft, Kim McLeod, Pepi
Fabbiano and Avi Loeb for valuable discussions. We also thank the
anonymous referee for useful comments. This work was supported by NASA
Contract NAS8-39073 (ASC) and NASA grant NAG5-3066 (ADP).


\begin{references}

\reference{b1} Bahcall, J.N, Kirhakos, S., Saxe, D.H., \& Schneider,
D.P., 1996, Ap.J. in press

\reference{b2}Boyle, B.J. et al., 1991, in Crampton D., ed. ASP Conf
Ser.No. 21, The Space Distribution of Quasars. Astron. Soc. Pac.,
San Francisco, p.191

\reference{c1}Caditz, D.M., Petrosian, V. \& Wandel, A., 1991,
Ap.J.Lett. 372, L63

\reference{c2}Cannizzo J.K. 1993, in ``Accretion Disks in Compact
Stellar Systems'', ed. J.C. Wheeler (World Scientific: Singapore), p.6

\reference{c3}Cavaliere, A., Giallongo E., Messina, A., Vagnetti, F.,
1983, Ap.J., 269, 57

\reference{c5}Cavaliere, A. \& Szalay, A.S., 1986, Ap.J. 311, 589

\reference{c6}Cavaliere, A. \& Padovani, P., 1988, Ap.J.Lett., 333, L33

\reference{c7}Cavaliere, A. \& Padovani, P., 1989, Ap.J.Lett. 340, L5

\reference{c8}Clarke, C. J., 1989, MNRAS, 235, 881

\reference{c9}Clarke, C. J. \& Shields, G.A. 1989. Ap.J. 338, 32

\reference{d1}Della Ceca, R., Zamorani, G., Maccacaro, T., Wolter, A.,
Griffiths, R., Stocke, J.T., Setti, G., 1994, Ap.J. 430, 533.

\reference{f1}Faber,  S.M., et al, 1996 preprint

\reference{f2}Fiore, F. \& Elvis,M. 1994, in: Proceedings of the COSPAR
Meeting, Hamburg, Germany, July 1994, in press

\reference{g1}Green, R.F., 1986, in: Quasars, IAU Symposium No 119,
Ed. G.Swarup \& V.K. Kapahi, Reidel, p. 429

\reference{h1}Haehnelt, M.G. \& Rees, M.J., 1993, MNRAS, 263, 168

\reference{l1}Lin, D.N.C. \& Shields, G.A. 1986, Ap.J.,305, 28.

\reference{l2}Lynden-Bell 1969, Nature, 223, 690

\reference{m1}Maccacaro, T., Della Ceca,R., Gioia,I.M., Stocke, J.T., Wolter,
A. 1992, in MPI fuer Extraterrestrische Physik:
``X-ray Emission from AGN and Cosmic X-ray Background'' p.291

\reference{m2}McLeod, K.K. \& Rieke, G.H., 1995a, Ap.J.Lett. 454, L77

\reference{m3}McLeod, K.K. \& Rieke, G.H., 1995b, Ap.J., 441, 96

\reference{m4}Meyer, F. \& Meyer--Hoffmeister, E., 1982, Astr. Ap., 106, 34.

\reference{m5}Mineshige, S. \& Shields, G.A. 1990., Ap.J. 351, 47

\reference{n1}Narayan, R. \& Yi, I., 1994, Ap.J. 428, L13

\reference{n2}Narayan, R., McClintock, J.E., Yi, I., 1996, Ap.J., 457, 821

\reference{p1}Press, W.H.,  \& Schechter,P.L., 1974, Ap.J., 187, 425

\reference{r1}Rees, M.Y. 1984, Ann. Rev. Astr. Astroph.,  22, 471

\reference{s1}Shakura, N.I. \& Sunyaev, R.A., 1973., A\&A, 24, 337

\reference{s2}Siemiginowska, A., Czerny, B. \& Kostyunin, V., 1996,
Ap.J. 458, 491 (SCK96)

\reference{s3}Smak, J.I. 1982., Acta Astr., 32, 199

\reference{s4}Small, T.A. \& Blandford, R.D., 1992, MNRAS, 259, 725

\reference{t1}Tanaka, Y., \& Lewin, W.H.G., 1995, in ``X-ray Binaries'',
eds. Lewin, W.H.,G., van Paradijs, J. \& van den Heuvel, E.P.J.,
Cambridge University Press, p.126

\reference{y1}Yi, I., 1996, Ap.J. 473, 645

\end{references}
\end{document}